# System Analysis and Design for integrated sponsored SMS/USSD Based M-Services

## A case study of Maternal Health M-Service in Tanzania

Timothy Y. Wikedzi[1], Ramadhani S. Sinde[2]
Computational and Communication Sci & Eng
Nelson Mandela African Institution of Sci & Tech
Arusha, Tanzania
[1]twikedzi@gmail.com, [2]ramadhani.sinde@nm-aist.ac.tz

Dan K. McIntyre [1]
Information Technology
University of Iringa
Iringa, Tanzania
[1] dkmcintyre1@gmail.com

*Abstract--*Mobile phones have proven to be the best way of providing reliable access to information to people in low and mid income countries where other forms of communication perform poorly. As a result of the wide spread of mobile phones, there has been an increase in number of Mobile Application (M-Services) which are being used as a tool for disseminating different type information to people. M-Services of this nature are established to address informational challenges that are faced by people especially low income people. Because of this then, these projects must be sustained so that people can enjoy the benefits of it. Contrary to this, reports show that most of these M-Services are facing the challenge of cost of operating them, which in a direct way affects the sustainability of these services. In this paper therefore we present an analysis and later design of a noncommercial M-Service, which integrates advertising functionality as a tool for subsidizing the cost of operating M-Services. To achieve this we have employed some concepts of Information System Analysis and Design (ISAD) as the guiding principle towards achieving our design. A prototype of M-Health is used for the study.

## I. INTRODUCTION

The mobile communications technology has quickly become the world's most common way of transmitting voice, data, and services in the developing world. They carry a potential of being the best media for dissemination of information because mobile services are widely available and inexpensive, [1]. Mobile phones are less inhibited by traditional access barriers that hinder the widespread use of many other communications technologies including geography, socioeconomic status, infrastructure such as electricity and literacy, [2]

The potential of Mobile Technology being the right tool for disseminating information has led to establishment M-

Service projects that are initiated to disseminate information to various social and economic groups within the society. An example of such projects operating in Tanzania and internationally are Mobile Alliance for Maternal Action (MAMA), M-Health Alliance, e-agriculture and Maji Matone and many other applications for Farmers, Sports, Educations etc. Existence of these projects have been reported to Improve livelihood of people by providing them with information which is an important tool for making informed decisions and in staying updated, [3]. [4] also reported that M-Services can contribute in fighting poverty by facilitating the convergence of local and global knowledge and disseminate it to the rural areas so as to improve economic production capacity in the settings in which the majority of the poor live.

Despite the fact that M-Service projects have proven to be a potential way for disseminating information to people, they are challenged by cost of service which is directly affecting sustainability of such services (MAMA 2013). mHealth Alliance report affirms to the fact that M-Services improves access to information, But the question of financial sustainability and ultimately "Who pays?" poses persistent challenges and barriers to scale and investment in such projects. Studies show that cost of information is among the major barriers to effective use of information which has resulted in underperformance and in some cases, total failure of M-Service e.g. Maji Matone Project in Tanzania. The experience of MAMA is a similar case for most other M-Services even those which are not related to M-Health. A report by mHealth Alliance on usage of mobile phone applications for disseminating information indicated that, at present, the majority of M-Services, particularly in low and middle-income countries, are dependent on donor funding. The report states further that, this model of financing is unsustainable because of the lack of certainty that funding will be renewed and always the funds are limited to allow the project to run in full scale. As a result, most of these M-Service projects do not survive because of their dependence on this form of financing. The question, who pays? Is at the centre of most M-Services discussion.

According to [5] report, Securing sufficient revenue is still a challenge for most providers of noncommercial M-Services. [5] Presents a solution that, M-Service providers must develop a creative mix of revenue streams while taking into account the affordability of services. The report identifies the challenge of limited income for people in rural areas and so recommends an advertising model to be used to cover the cost of operating the M-Service.

There are M-Services which have succeeded and based on the current trend they have shown potential for further adoption (intermedia 2013), examples of such services are Mobile Money (M-Money) and Mobile Banking. Which uses a self-financing mechanism by charging users per transaction. There are also existing commercial M-Services agencies which run mobile services for information access such as Health information and agricultural information on weather and market prices for farmers as well as other forms of SMS campaigns, [6], [7]. Instead of providing free access to information these agents charge all their subscribers an access fee that ranges between 250Tsh, 500Tsh or higher for access. These models of financing M-Services promise sustainability but they pose a threat of Digital Divide between those who can pay for access and those who cannot pay. Cost of service in running M-Service has direct impact to the sustainability of M-Services. Donor funded M-Service project will only survive as long as the project gets support, but as the size of the project grows it then becomes a direct threat to sustainability of the M-Service. A sustainable financial model is needed to ensure survival of M-Service projects [8]

This study is set to adopt the recommendation from [5] and design an M-Service that integrates Advertisements functionality as system module for subsiding cost of operating an M-Service. Advertisements have widely been used in the field of Information technology as a mechanism for covering the cost of access and of course making profits. Companies like Google, for example, make a fortune through advertisements. The potential of advertisements in covering the cost of access has benefited a lot of other companies like Facebook and Yahoo which provide users with free access at the cost of receiving advertisements. Nowadays there are thousands of free mobile apps in the market which use advertisements as a tool for generating income. However in this study we are set to present an idea of SMS advertisements as a potential solution for free access to information for all categories of mobile phone users.

Implementing a sustainable M-Services is an important aspect towards ensuring long term benefits of those projects. In this paper we present the Analysis and Design of an Integrated M-Service that addresses the challenge of cost of operation by using SMS advertisement support as an integral part of the system. A case study of M-Service for Maternal Heath in Tanzania is used.

## II. LITERATURE REVIEW

### A. Mobile Application as M-Service

[5] Defines mobile Applications as software designed to take advantage of mobile technology. In this paper we specifically refer to Mobile Applications as SMS Based Mobile Application (M-Services). This way we put a distinction of M-Service from more complex and advanced Mobile Applications that run on smartphones. The reason to do this is that, although mobile phones are widely spread in low and mid income countries, the majority of people, especially in low income countries, own basic phones which have limited features compared to smartphones.

### B. Sustainable M-Service

[9] Defines Sustainability as the process of maintaining something that already exists over time without needing an outside support for it to continue existing. Whereas [10] provides us with a more specific definition of sustainable IT as a Technology that is capable of being maintained over a long span of time without being affected by the changes in both hardware and software. This definition presents to us two important things to consider when talking about sustainable IT, in this case an M-Service, one is the life span of the system, and two is the ability of a system not being affected by the changes in hardware and the software.

### C. SMS

Short Message Service (SMS) is a text messaging service component of phone, Web, or mobile communication systems. It uses standardized communications protocols to allow fixed line or mobile phone devices to exchange short text messages, [11]. Content of one SMS is limited to a maximum160 ASCII characters. SMS are sent to mobile phones via the SMS Gateway. Using SMS technology, it is also possible to send them in Bulk, in which one or more SMS are sent to more than one user. This offers a better way of reaching out to more people at a time.

### D. Ad

This is a term that is used to refer to Advertisement. In this context, it always refers to a text base advertisement that is intended to be sent to the end user inform of an SMS.

### E. SMS advertising

SMS advertising is a subset of Mobile advertising, a form of advertising via mobile phones or other mobile devices. Other forms of mobile advertising which are specifically common in smart mobile devices are Mobile Web Banner (top of page) and Mobile Web Poster [12]. The focus of this paper though is on SMS advertising, which has been reported to be the leading form of Mobile advertising worldwide, for the reason that our proposed system is also for no smartphone users.

### F. Approaches of pushing SMS advertisements

1) *Receiver Opt-in:* This form of SMS advertisement involves requesting for the consent of the potential receiver before starting pushing of advertisements.

2) *Purchase of Receiver numbers from third party source:* This form of advertisement involves purchasing of receivers contacts from third party companies and using them to push SMS advertisements

### G. USSD

[13] Defines *Unstructured Supplementary Service Data (USSD)* as a communication protocol used to send text messages between a mobile phone and applications running on the network. It is a messaging service used in Global System for Mobile Communications (GSM) networks similar to SMS. However, unlike SMS; USSD provides session-based connections. Because of its real-time and instant messaging service capability, USSD service offers better performance and is much cheaper than SMS for two-way transactions. This service is unique and only available to GSM networks. The following are the advantages of USSD as described by [13].

1) *USSD code format*

USSD communication is initiated by dialing a special code. USSD codes comprise of asterisk (*), followed by a combination of digits (0 to 9) and a hash (#) Example *150*00#. The * and # codes are used to signify the beginning and end of the request.

### 2) USSD Architecture

According to [13] The USSD architecture basically comprises

- The network part that includes the Home Location Register (HLR), Visitor Location Register (VLR), and MSC

- Simple Messaging Peer-Peer (SMPP) interface for applications to enable services

- USSD Gateway and all specific USSD application servers

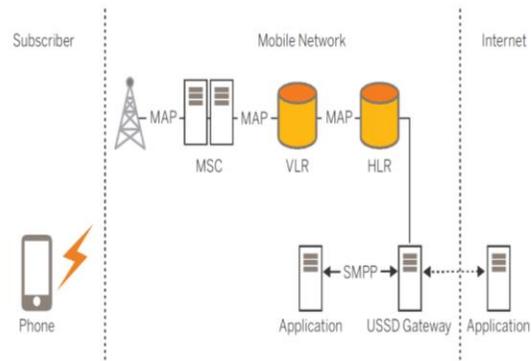

Figure 1: Elements of the USSD Mobile Network (Source Aricent)

While USSD seems to be a complex architecture, the focus of this study is not in analyzing USSD architecture but rather to make use of it as a black box. Our only concern is on the Internet and Subscriber sides. In this paper we present the analysis and design of an Application that that interfaces with USSD to reach the end users.

## III. OVERVIEW OF THE SYSTEM

In this paper, we present the analysis and Design of an M-Service (m-Health) that rely on sponsorship (advertisement) to convey maternal health information to the public of Tanzania. The proposed system will involve the use of mobile phones to deliver messages to the target audience. The system will allow the user to subscribe and get information via short messages. All messages stored on the system will be provided by health professionals to ensure that the information sent to users are correct. These health professionals will be assigned special accounts that they will use to access and work in the system. End users will be able to retrieve this information via SMS using their mobile phones. They will also have the opportunity to send various questions and receive answers from medical professionals. The system is designed with a capability of sending advertisements to end users as a way of covering the cost of operation. To ensure future sustainability of this system, the system design also integrates a payment system for economic buyers who will not want to receive advertisements. Payment service will also be useful in time if there are no funds to cover the cost of free access and in meeting the needs of those who would want unlimited access. However, the focus of our analysis and design is on Sponsored access (access through advertisements).

### A. Integrating Technologies

The Architectural design of our M-Health is based on four tiers architecture (see **Error! Reference source not found.** ) as it has been proposed by [14]. They identified that using a four tier architecture for mobile applications offers more abstraction, more independence between components of the system, and hence a more flexible way of implementing mobile applications. We have adopted this architecture because we see it fitting our design idea. It also gives us a great opportunity to plan for growth as mobile technologies evolve so fast and the future holds that more mobile phone users will be upgrading to smartphones and other devices.

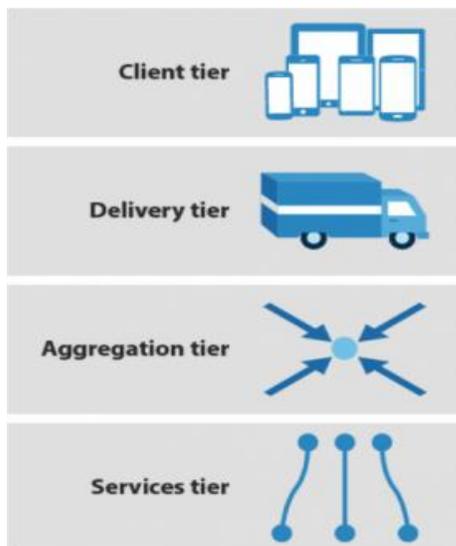

Figure 2: Adopted Four Tier Architecture (Source: http://blogs.forrester.com/)

### B. Implementing the Four Tier architecture for our M-Health

- *A client tier*: Represents end users means of accessing our M-Health System. The client tier in this case refers to both mobile phones and web clients. This presentation layer creates a separation between the applications and accessing devices and the backend services that the application delivers.

- *Delivery tier*: Is the channeling tier. It handles the physical routing of requests and information to

and from the system. This is the direct interface of the systems and the end users. In our M-Health Systems this represents Telecommunication channels and services such as USSD and SMS. It also represents the Internet as the channel through which Medical Experts will be interacting with the system.

- *Aggregation tier*: Acts as a bridge between data and upper tiers. It provides means for handling all data requests and exchanging data between service tier and the Delivery tier. Our M-Health System will run on PHP as the language for our application.

- *A services tier*: Handles all data functionalities. It includes the database that stores and manages all data. The choice of Database for our M-Health system is a MySQL database. The reason for choosing this type of Database management system is that it's open source and has a strong community for support.

### C. Sponsored (Advertisement) access Request life cycle

Under this subsection we describe the concept of the advertisement mechanism for our proposed M-Service. Of the two approaches for pushing advertisements to the end user, we will be using the opt-in approach. Therefore, all our designs are based on an assumption that the end user opted to receive an advertisement

#### 1) Information Requesting

Information processing begins with the user, sending a USSD request from his mobile phone. After dialing a special USSD code, the user receives a USSD menu from which he can choose categories of information to access. Information request is limited to categories after which a random information will be sent to the end user base on the selected category.

#### USSD menu generation

The USSD menu is dynamically generated from the defined categories of information in the database. This menu gives a user access to various types of information that is stored in the database. Generating menus dynamically adds more flexibility to the system.

#### 2) Advertisement processing
##### a. Fetching of an Ad from the database

All advertisements will be stored in the database and associated with sponsors who will be registered into the system. The process of fetching an ad from the database is triggered internally after the M-Service has received an access request. Before an ad is sent, the system checks to see if there are any existing ads. The conditions for an ad to exist is that there should be at least one sponsor with some remaining balance. This is how the system will know that a message can be sent. Once the ad has been confirmed to exist the system will proceed on fetching the current ad in the queue and submit it for further processing

### b. Processing ads based on available sponsors.

In this initial design, all ads will be sent in a queue that will rotate around all available sponsors whose balance is above zero. The future design of this system will allow ads to be sent to a client base on the relevance of the locality as the first priority. This way the system will be able to send context based and relevant ads to end users. This of course will be achieved with an upgrade to smartphones.

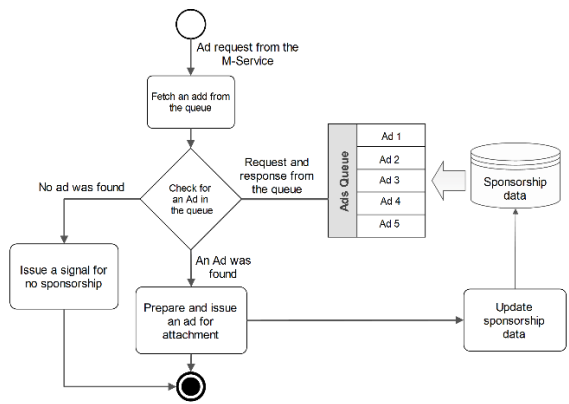

Figure 3: Design concept of the advertising mechanism

### 3) Fetching of information

Once sponsorship has been confirmed, content is then prepared to be sent to the users. As it has been stated earlier, information to be sent to the user will depend on the category of information that was picked by the user.

### 4) Integrating an ad with requested information

The next step after fetching both the ad and the information is to bind them together and send them as a single data packet to the user. There are two alternatives on how this information will be sent to the user; first, we send two separate SMS one for the ad and the other with the requested information. Second, we send one SMS containing both the ad and the content.

### 5) Content Delivery

Though all information requests will be issued via the USSD service, the requested information will be sent back to the user via an SMS. This is because USSD messages are session based, so once the session is over the returned message is lost. Also USSD messages cannot be saved in the inbox of the end user's phone.

## IV. ISAD FOR SUSTAINABLE M-SERVICE

In Information Systems (IS) development, Analysis and Designing is used as a tool for developing successful IS. History and experience prove that most of the IS projects that had in one way or another ignored ISAD suffered great failures in the form of exceeding budgets, failure to meet the running cost of the system and in some cases total rejection of the system by the users. ISAD as a tool is a part of System Development Life Cycle (SDLC). Dennis, A. et el, 2012 define the Systems development Lifecycle (SDLC) as the process of determining how an information system (IS) can support business needs, designing the system, building it, and delivering it to users. The key concept in this definition is that IS are need based, meaning that there is no purposeless IS as they are all developed for the sake of addressing a certain need. In this paper, we present ISAD considerations for sustainable M-Service. We present the analysis and Design of an M-Service that integrated Advertisement as a tool for subsidizing the cost.

### A. Identification of end users

As a part of ISAD, the project must clearly identify the end user, in the case of M-Service the end user is referred to as the individual(s) that and are intended to be accessing information offered by the service. The process of identifying users of an M-Service is highly dependent on the type of M-Service. If for example a service is meant for Farming activities, then clearly it means that farmers form a primary group of end users of that service. In the case of an M-Health (Maternal Healthy) system which we refer to as a case study in this paper, the end users of the system are primarily Pregnant women, Fathers, Mothers and all those who would benefit from accessing maternal health information. User identification as an aspect of ISAD is necessary to ensure that a solution based M-Service is developed and better design is later achieved.

### B. Considering the cost of operation

Well, this is the focus of this study. As it has been reported that covering the cost of operation is a challenge that most M-Services face,. The cost of operation is a critical factor to be considered in any M-Service since they all rely on Mobile communications which are run commercially. Mobile network operators dominate the M-Services ecosystem in developing countries. They serve as gatekeepers, and dictating matters related to revenues including requiring payments for the service [5].

M-Services are categorized into two major types; commercial M-Services which are typically developed to deliver information or conduct transactions (or both) and noncommercial M-Services which are developed only to provide information [5]. Commercial M-Services require end users to pass through some form of payment when accessing the service. Sponsored (noncommercial) M-Services on the other hand have a third person who covers the cost so that end users can enjoy a free service. Running a Commercial M-service is clear when it comes to costs of operating them. Because end users are charged for the service and in so doing the host of such application makes a profit and covers the cost of operations. Examples of such M-Services are M-Banking, M-Money etc. In Tanzania the leading Commercial M-Service provider is Vodacom with their M-Pesa service which is a form of M-Money service that offers a service of mobile money transactions.

The Analysis and Design focus of this paper is on Sponsored M-services because, they have potential of or extending information accessibility to people of low income. Considering the current economic state of people in rural areas of Tanzania which according to census results of 2012 more than 60% of the whole population live in rural area where there is extreme poverty.

### C. System and User interactions

The question of how users will interact with the system is critical in any IS project because it determines the type of technologies to be used as well as in planning for the cost of operation. Our M-Heath System offers two types of users Web Client users (Administrator and the Medical Experts) and Mobile subscribers who are the end users of our M-Health System. Web users interact with the system through Web browsers and internet whereas all our mobile subscribers uses their mobile phones to access the system. The mode of communication between the System and the subscribers will be via SMS/USSD services. Subscribers will also be able to post questions by sending them via SMS to a special number. Administrator and Medical Experts will be able to add new content and manage the system from the backend.

### D. Requirements gathering and Analysis

[15] Defines Requirements as statements of what the system must do or what characteristics it needs to have. Any IS relies on requirements for its implementation an M-Service is no exception. However M-services present a challenge when it comes to gathering of requirements, M-Services are IS that are not domain based. They are

developed to be used as a medium of disseminating information to open end users, it could be a well-defined group of end users, like in this case pregnant women for example. But still these people are not organized under one domain, they are all independent and with freedom to choose to adopt the service or to remain as watchers. We understand though, the idea of our M-Service is that it will be delivering simple text messages either in form of SMS service or USSD messages. End users of our application need not to know how the backend of our application is structured. Our Analysis and Design is therefore characterized by the idea of a information system that delivers information to end users in a manner that users can afford to get it, access it easily, and understand the content.

*1) M-Health application requirement*

Since the focus of this paper has been on Sponsored M-Services we will then discuss requirements based on that sector. Before defining requirements we need to describe how it will work with use cases. In a domain limited IS projects, requirements are derived from the business processes of a particular domain e.g. Human Resource IS, Academic IS, Payroll IS etc.IN the other hand, M-Services are open ended systems, meaning that the developer proposes the requirement and implements the system. The role of end users is not to provide initial requirements but rather to give feedback that will later improve their experience with the service.

*2) Functional Requirements*

Functional requirements are those requirements that are used to illustrate the internal working nature of the system. They describe what tasks the system should perform.

*a) Subscribers*

- Subscribes
- Send questions
- Access Information via mobile phone

*b) Administrator*
- Manage Doctors
- Monitor Subscribers
- Manage Sponsor Information
- Access Information

*3) Nonfunctional Requirements*
*a) Operational*
- USSD/SMS service

- The system should interface with SQL database.

b) *Maintainability & Upgrades*

- System should allow upgrade to smartphone usage

- System should allow updates and upgrades without affecting users' experience

c) *Acceptance*

- System Language should be Swahili

These are a summary of all requirements that as a developing side we have come up with. It is of course true that some or most of these requirements will change overtime as users will require some aspect of the systems to be improved, added, or dropped. There is at all-times a call for better future versions of the system and services.

*E. Structuring of the Requirement*

Having defined the requirement for our M-Service we move to the next stage of systems analysis. According to [16] system analysis phase consists of two parts, determining requirements and structuring requirements. During structuring of the requirements the goal is to interpret and model processes and data for our application. A common tool used for structuring information is the Data-Flow Diagram (DFD). [16] Describes **data flow** as data that are in motion and moving as a unit from one place in a system to another. DFDs use four symbols, data flows, data stores (e.g. database), processes, and sources/sinks (or external entities) to represent both physical and logical information systems.

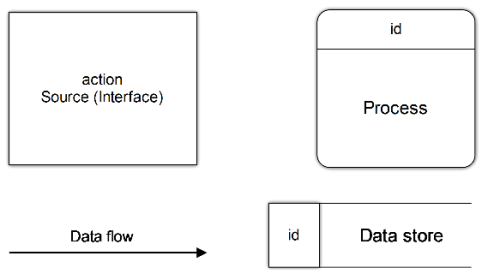

Figure 4 : DFS tools

1) *Process flow modelling:* Process modelling is meant to provide a description of how data flows from one point to another. Our M-Health system has three possible users who will act as the action source to our system. These users interact with the system with different roles and each causes different types of data to flow.

a) *Context Diagram:* A Context Diagram in this case is meant to give a general concept of the system and

data that will be flowing as ads to the identified interfaces (users). Figure 5 Shows the context diagram for the proposed System. It consists of the main process Health SBMA (SMS Based Mobile Application)

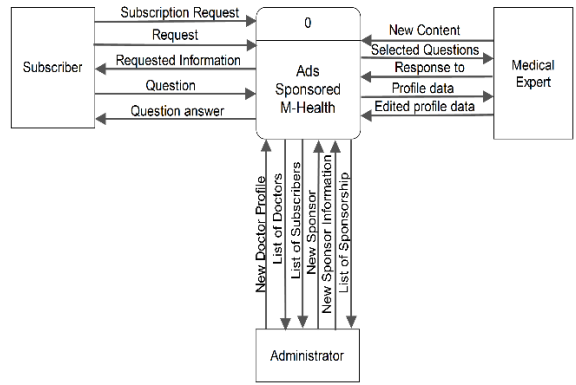

Figure 5: Context Diagram showing general data flows for our mHealth

b) *Level 0 Process Diagram*

Expounding the context diagram, we create the object Level 0 diagram. The level 0 diagram basically breaks down the Health SBMA (SMS Based Mobile Application) as identified in the context diagram. Figure 6 show the diagram for level 0 of the proposed M-Health System.

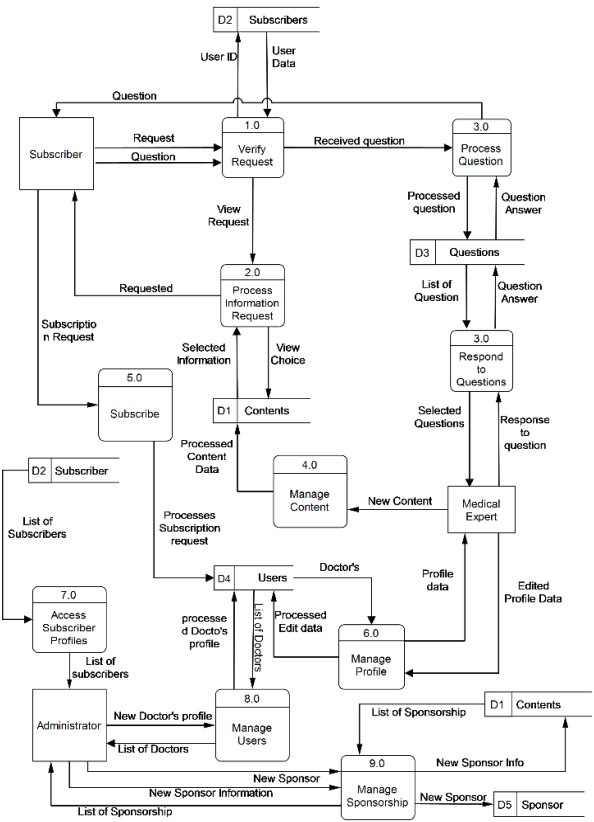

Figure 6: Level 0 data flow diagram describing processes in more details

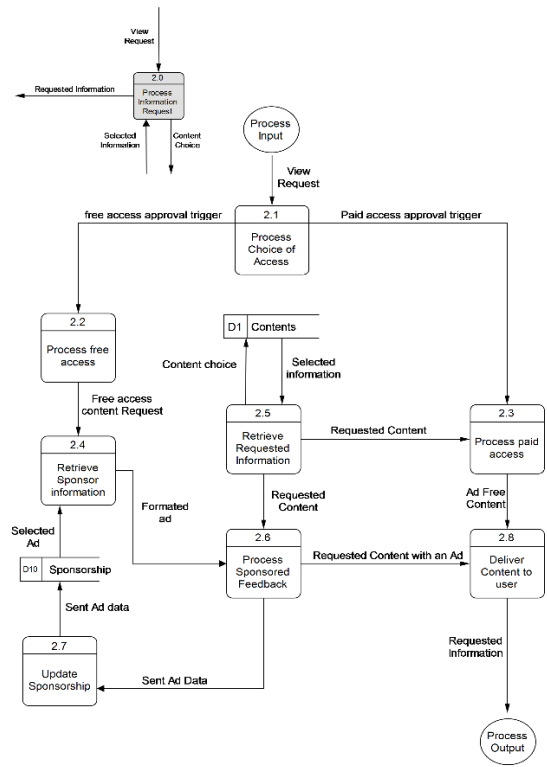

Figure 7: Level 1 Data low diagram for Process 2.0 in Level 0

### c) Level 1 Process Diagram

Level 1 diagram for our M-Health describes the breaking down of process 2.0 which handles the task of processing information requests. There are two types of request that the system can receive, one is of free access to information and the other one is for the paid access.

### 2) Conceptual Data Modelling

[16] describes conceptual data modelling as a way of representing organizational (Information Systems) data. The goal of data modelling is to show as many rules about the meaning and interrelationships among data as possible, independent of any database management system (e,g MySQL, Oracle, MS SQL, SQLite). The common tool that is used to model the data are Entity-relationship (E-R) data models. These are diagrams that show how data are organized in an information system. ERD uses a special notation of Rectangles (to describe entities), diamonds (to describe relations), and lines to represent as much meaning about data as possible. The deliverable from the conceptual data-modeling step within the analysis phase is an ERD Diagram

During the analysis of requirements for our M-Health we had put focus on data to gain the perspective on data needed to develop a data model. The following entities have been identified; (1) USER_GROUP (Defines roles of users e.g Doctors, Administrator), (2) USER (Handles users' profile data), (3) SUBSCRIBER (Stores subscribers information) , (4) QUESTION (Stores posted questions), (5) ANSWER (stores responses to questions), (6) CATEGORY (Stores categories of information which forms menus and submenus), (7) CONTENT (Stores information that subscribers can access), (8) RECEIVED_SMS, (9) SPONSOR (Stores sponsors information), (10) ADS (Stores all the advertisements)

The conceptual data model of these schemas is represented diagrammatically by Figure 8

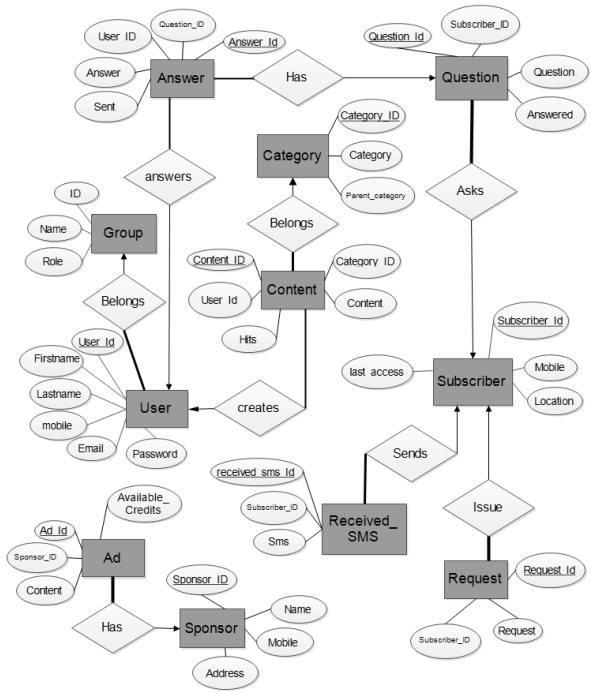

Figure 8: ERD Diagrams describing the conceptual model of our mHealth System

The ERD diagrams presented in this subsection forms the basis for the database implementation.

## F. System Designing Considerations

### 1) Keeping it simple

SMS based application are characterized by the size of the SMS which is determined by a fixed number of text characters. The size of SMS directly affects the cost of operation because the cost of sending SMS increases with the number of Messages. For this reason, then it is important that the overall application is kept simple and more focus to be put on means for optimizing the amount of information for an SMS.

### 2) Information Design

[17] States that a high quality information design communicates information in a manner appropriate and pertinent to a reader's situational context. It must focus on the reader ability to understand it and to extract meaning from that information. Information Design is one of the very important factors for a sustainable M-Service. Our

M-Health System is meant for Tanzanians and so information design aspect should consider the nature of these people to determine the type of information and in which form that information going to be delivered. The national Language of Tanzania is Kiswahili? And it is both the official language and the language that most people understand. It is obvious then, one of information design criterial for our M-Service is that its content must be in Swahili. Another very important factor to consider is, as has been mentioned before, that this application relies on SMS/USSD services. The cost of these services is determined by the size of message

### 3) Designing Interactive menus for USSD

All end users (Subscribers) of the system will access the system via their mobile phones. The system is designed to work for all types GSM of phones. **Error! Reference source not found.** Shows a summary of user actions when accessing our M-Health application via the mobile phones. The flow of actions is numbered from 1 to 12. (1) User enters the USSD code to access information in this case it is *31022, (2) If the user is accessing the system for the time he will be asked to subscribe by sending a messages to a special number (registration could free or onetime payment of 250Tsh). (3) Once in the system the user will receive a USSD message informing him that the service is free but he will receive an ad. (4) User can then choose to proceed or quit. (5) Upon agreeing to continue a menu of information categories will be presented.(7) More submenus will be presented for user to narrow down his choice. (9) After reaching to the last subcategory, user will receive a USSD SMS informing him that information will be sent to his phone as an SMS. (10) User will then receive an advertisement with a code that he will have to send to the system via USSD. Sending of this code is a way of making sure that at least the user read the Ad SMS. (12) Lastly the requested information

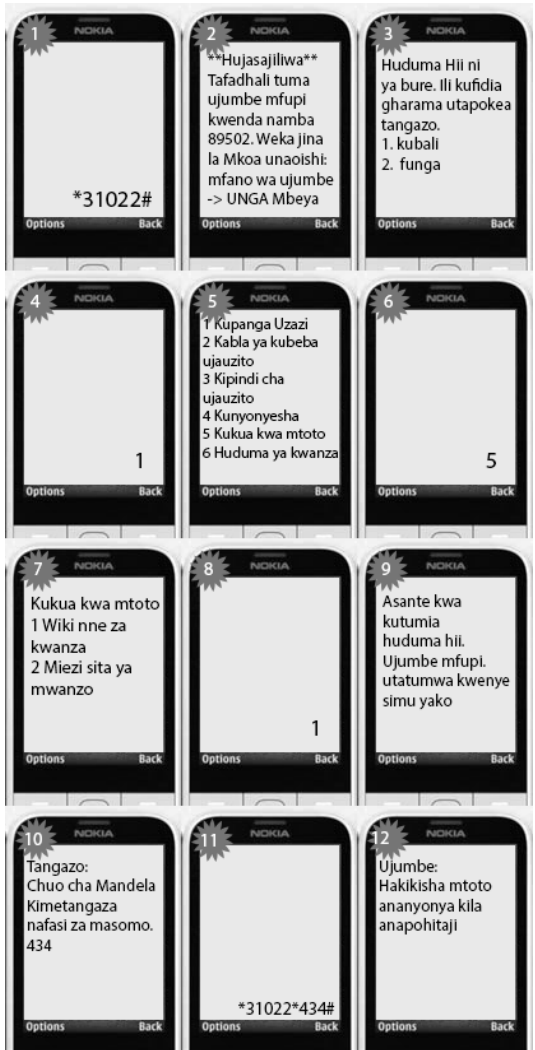

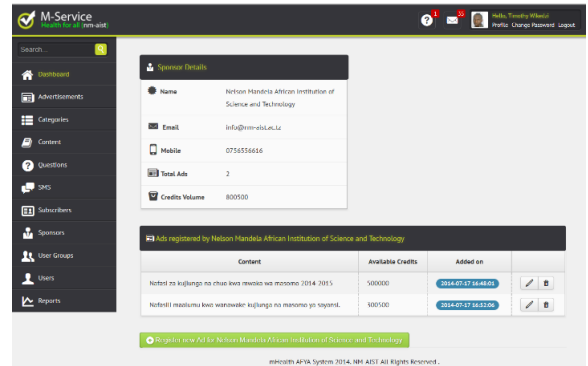

Figure 10: Sample web Client page interface for administrators and Medical Experts to access the system.

### G. Ads Consideration

In the context of this paper, we strictly consider advertisements as an information and therefore the concepts of information design also apply to them. It is important for all the advertisements to be categorised so that right ads can be delivered to the right people. Ads can be delivered base on Locality of an individual but also base on the type of content that the user requested. Some ads may be useless to people who cannot afford to pay for what is advertised, Though not each of our subscribers will be poor, it is still important to determine the type of advertisments to be sent to subscribers.

### H. Defining the right source of information

Sustainable M-Service must have a reliable source of information. Reliability of information in this case does not mean only availability of such information, but it also implies trusted sources of right information. We do not want to feed our users with false information this could easily lead to misleading and loosing trust of our users. Our M-Health service will rely on Qualified Doctors and other Medical practitioners as the source of content for our M-Health.

### I. Conclusion

In this study, we set out to do a case study on how to develop a system which is maintainable and sustainable in a community. By reflecting on the process, and the system being designed we intended to draw conclusions of the important aspects which exists when developing such a system. Our study is based on a M-Health prototype application that was analyzed and designed based on the knowledge of how existing M-Services work from end users perspective. This way we were able to design our M-Health prototype which will later be improved through feedback from end users.

We acknowledge the fact that a sustainable IS puts into consideration the growth aspect. It is then meaningless for

Figure 9: Action Flow UI's whowing end Users Usage of the SYstem (Content on phone is in Swahili because the targeted end users are the Tanzanians)

#### 4) Web interface Design

Both system administrator and the doctors will access the systems by using a web interface. This way they will be able to add and manage new users and manage new content respectively.

a system to be developed to offer a useful service or solution and due to failure of shifting to new technology. This being the case, we recommend that a sustainable IS should not be tightly coupled with technology but rather should run independently of it and be flexible to change and adopt new technologies whenever a change is required, example changing to smartphones and modern technologies that are supported by mobile devices. SMS Based M-Services must therefore be developed strategically ready to be upgraded into smartphone applications, this will guarantee greater chances of continuing provision of the service. We have chosen to adopt the four tier architecture that decouples Data, processing application, Distribution Channels and Client technologies. This way our application can integrate and work well with Mobile devices, Web Clients, Any Data source and can potentially be distributed via internet and GSM networks.

As an outcome of this study and analysis, we finally conclude by acknowledging that M-Services have proven to possess a great potential of addressing information challenges that are facing low and mid income countries, therefore they must be kept sustainable to ensure that the intended service is continually offered for the benefit of people.

## RFERENCES

## AUTHORS PROFILES

TImothy Wikedzi, is a Master's student at the Nelson Mandela African Institution of Science and Technology. He is pursuing a master's degree in Communication Science and Technology. He currently lives and study in Tanzania

Dan K. McIntyre, is an Adjunct Professor at University of Iringa in Tanzania. He is a an expert in Computer Science particularly Software Engineering. He currently lives in United States of America.

Ramadhani S. Sinde is an Assistant lecture at Nelson Mandela African Institution of Science and Technology. He is an expert in Telecommunication.